\documentclass[preprint,3p,12pt, authoryear
]{elsarticle}

\journal{Preventive Veterinary Medicine}
\usepackage[T1]{fontenc}
\usepackage{mathtools}

\usepackage{booktabs}
\usepackage{array}
\usepackage{tabularx}
\usepackage{longtable}
\newcolumntype{L}[1]{>{\raggedright\arraybackslash}p{#1}}
\usepackage{threeparttable} 

\usepackage{algorithm}
\usepackage{algorithmic}

\usepackage{graphicx}
\usepackage{float}
\usepackage{pdflscape}
\usepackage{multirow}

\usepackage[symbol]{footmisc}
\usepackage{enumitem}
\usepackage{changepage}
\usepackage{xcolor}
\usepackage{placeins}
\usepackage{appendix}

\usepackage[hidelinks]{hyperref}

\biboptions{authoryear} 
\usepackage{booktabs}
\usepackage{tabularx}
\usepackage{siunitx} 
\usepackage{caption}

\begin{document}

\begin{frontmatter}

\title{Multi-network comparison of between-farm contacts for infectious disease surveillance in swine production}

\author[1]{Jason A. Galvis} 
\author[1]{Nicolas C. Cardenas} 
\author[1]{Gustavo Machado\corref{cor2}}
\ead{gmachad@ncsu.edu}

\cortext[cor2]{Corresponding Author.}

 \affiliation[1]{organization={Department of Population Health and Pathobiology},
             addressline={North Carolina State University}, 
             city={Raleigh},
             state={NC},
             country={USA}}
        
\begin{abstract}
Understanding how swine farms are interconnected, directly and indirectly, is essential to characterizing infectious disease transmission. This study aimed to describe the connectivity of swine farms across 11 network types, including vehicle movements (i.e., trucks and trailers), animal movements, and distance-based farm-to-farm contacts, to identify links among production types and farms likely to be consistently characterized as super-spreaders. Network edges represented vehicle movements, pig movements, and farm-to-farm contacts within a 10 km radius. Truck and trailer movement networks were the most densely connected, particularly for feed transport, showing connectivity levels between 98.7\% and 99.7\% higher than those of pig movement and distance-based networks. These networks also exhibited the highest degree and frequency of connections between farms, while the aggregated truck network, which included all truck types, showed the greatest potential to act as a bridge connecting farms. Finisher farms were highly interconnected with other farm types across all networks. Sow farms were frequently reached by other farm types, especially through feed truck movements, representing up to 8.7\% of these links. We demonstrated that in vehicle movements and proximity networks, finisher farms played a major role as super-spreaders. When comparing the top 50 farms ranked by super-spreader score in each network, vehicle-based networks showed the highest similarity, with up to 89\% of top-ranked farms shared between vehicle networks. In contrast, pig movement and distance-based networks identified largely distinct sets of top-ranked farms, sharing at most 4\% and 8\%, respectively, with other contact networks. Overall, each network exhibited a distinct connectivity structure, resulting in different sets of high-risk farms, particularly regarding potential transmission to breeding farms. These findings support the integration of multiple transmission pathways into disease surveillance.


\end{abstract}

\begin{keyword}
Sentinels, control actions, outbreak management, tracing.
\end{keyword}

\end{frontmatter}


\section*{Introduction}

Understanding how farms are connected through direct and indirect contacts is essential to characterize the potential for infectious disease transmission in livestock production systems \citep{lentz_disease_2016, machado_quantifying_2021, cardenas_spatio-temporal_2021, acosta_network_2022}, particularly in the U.S. swine industry due to its highly vertically integrated structure \citep{cardenas_analyzing_2024, passafaro_network_2020}. Traditionally, between-farm animal movements and spatial proximity have been considered the primary pathways for pathogen dissemination \citep{galvis_modelling_2022, galvis_modeling_2022, guinat_transmission_2016, olesen_transmission_2017}. However, growing evidence indicates that other indirect contacts, such as vehicle movements, can also play a critical and often underestimated role in the spread of diseases such as porcine reproductive and respiratory syndrome  (PRRS) \citep{galvis_modelling_2022, dee_experimental_2004}, porcine epidemic diarrhea (PED) \citep{galvis_modeling_2022, vanderwaal_role_2018}, and African swine fever (ASF) \citep{yoo_transmission_2021, gebhardt_sampling_2022}.

Although the ability of pathogens to remain viable on vehicle surfaces decreases over time \citep{mil-homens_assessment_2024, jacobs_stability_2010, quinonez-munoz_comparative_2024, pujols_survivability_2014, boniotti_porcine_2018}, the high frequency of feed and pig shipments, together with the availability of vehicle fleets, results in frequent movements with short intervals between farm visits \citep{galvis_role_2024, galvis_mitigating_2025, thakur_analysis_2016, masserdotti_role_2026}. Furthermore, while pathogens are expected to be inactivated through cleaning and disinfection (C\&D), the effectiveness of vehicle C\&D varies widely, with several studies reporting limited efficacy regardless of the disinfectant or C\&D method used \citep{parker_evaluation_2025, parker_evaluation_2025-1, boniotti_porcine_2018, houston_evaluation_2024, mannion_investigation_2008}. Under these conditions, vehicles can connect a large number of farms, as shown in previous studies \citep{galvis_role_2024, galvis_mitigating_2025}, and, when contaminated, may act as mechanical vectors across production systems \citep{yoo_transmission_2021, gebhardt_sampling_2022, parker_evaluation_2025, masserdotti_role_2026}. Given the growing concern about the role of vehicles in disease spread, several studies have used transmission models to evaluate their contribution, with results suggesting that vehicle-mediated contacts may account for a considerable proportion of transmission events \citep{galvis_modeling_2022, galvis_modelling_2022, sykes_estimating_2023, prezioso_network_2025}. These findings highlight the need to further investigate the role of vehicle-mediated transmission and better characterize the unique structural properties of vehicle contact networks, particularly when compared with networks based on animal movements and spatial proximity.

In regions with high pig density, transmission driven by close proximity becomes particularly relevant \citep{sanchez_spatiotemporal_2023}. This type of transmission is likely influenced by several factors, including farm ownership and the strictness of biosecurity protocols that influence the movement of people, equipment, and vehicles between farms \citep{fleming_enhancing_2026, alarcon_biosecurity_2021}. In addition, other mechanisms such as airborne spread, vectors, and wildlife can further facilitate transmission among geographically close farms \citep{guinat_transmission_2016, yoo_transmission_2021}. Distance between farms is typically measured using geodesic distance, defined as the shortest path between two locations on the Earth's surface \citep{hijmans_geosphere_2022}. However, for transmission pathways involving the movement of equipment and personnel, road distance may provide a more realistic measure of connectivity than a straight line distance between farms. Unlike geodesic distance, road distance reflects the actual routes used for transportation, which can substantially influence the likelihood of contact between farms \citep{cardenas_analyzing_2024}. Greater road distance may reduce the probability of interaction, even when farms are geographically close. Therefore, it is important to characterize how farms are connected through both geodesic and road-based distances to better understand potential transmission pathways.

It is generally assumed that the transmission of infectious diseases in swine production systems follows a vertical production flow, primarily driven by the movement of pigs among production types (e.g., sow → nursery → finisher) \citep{cardenas_analyzing_2024, passafaro_network_2020, lentz_disease_2016}. However, connectivity patterns between farms can differ substantially depending on the transmission route, including vehicle movements and spatial proximity \citep{prezioso_network_2025}. As a result, contact patterns across production phases, along with the identification of farms that may act as potential super-spreaders, are likely to vary across networks. Despite this, a comprehensive benchmarking comparison of multiple network types has not yet been described, which is essential for designing effective surveillance and control strategies.

Several studies have evaluated pig and vehicle movements, as well as the spatial distribution of farms, for disease surveillance purposes \citep{galvis_modeling_2022, galvis_modelling_2022, sykes_estimating_2023, andraud_modelling_2022, prezioso_network_2025, vanderwaal_role_2018}. However, few studies have simultaneously compared animal, vehicle, and distance network types or explicitly quantified proportional connectivity among farms and production types \citep{prezioso_network_2025, galvis_descriptive_2026}. Such analyses could identify high-risk transmission farms and pathways, and support targeted intervention strategies. Therefore, the objective of this study was to quantify proportional connectivity among farms and production types using multiple data sources of several commercial swine production systems in 18 U.S. states, including pig movements, truck and trailer movements, and geodesic and road-based distances. This approach provides a comprehensive understanding of how production systems are interconnected through both direct and indirect contacts.

\section*{Methodology}

\subsection*{Data}

We used RABapp\textsuperscript{\texttrademark} \citep{fleming_enhancing_2026} data, of which we collected information from 3615 farms, including farm identification, geographic coordinates, animal capacity, and production type, from four swine companies with production in 18 U.S. states. In addition, we collected the barn layout for each farm using the line of separation (LOS) component of premises biosecurity plans, as available in the RABapp\textsuperscript{\texttrademark} biosecurity mapping component \citep{fleming_enhancing_2026}. We also collected centroid coordinates for 51 cleaning and disinfection vehicle stations, 32 slaughterhouses, and 55 feed mills used by these swine companies, and processed them to delineate the perimeter of each location. From the four companies, we collected 79997 pig movement records, including source, destination, number of pigs transported, and dates, from January 1st, 2024, to December 30th, 2025. In addition, we collected 1.2 billion GPS records of truck cabs (hereafter referred to as trucks) and 297 million of trailer movements from January 1st, 2024, to December 30th, 2025. These data consisted of latitude and longitude coordinates recorded every five seconds, along with vehicle identification, date, time, and speed \citep{galvis_role_2024}. For trucks, the ambient temperature recorded by the GPS added sensors was also available. To complement trailer data, we obtained raster temperature data from Google Earth Engine at a spatial resolution of 1 km$^2$ \citep{abatzoglou_development_2013}, which was used to assign environmental temperature values to trailer locations during movements \citep{galvis_role_2024}. We also collected metadata describing the fleet of 1634 trucks and 2166 trailers, including vehicle identification and working role: feed (vehicles assigned to feed transport), pig (vehicles assigned to pig transport between farms), and market (vehicles assigned to pig transport to slaughterhouses).

For 52\% of trucks and 57\% of trailers, the working role was not available. To address this, we inferred the vehicle's working role based on its visitation patterns. Specifically, vehicles with a higher proportion of visits to feed mills than slaughterhouses and C\&D stations were classified as feed. Vehicles with a higher proportion of visits to slaughterhouses and C\&D stations than feed mills were classified as market. Vehicles with no visits to both feed mills and slaughterhouses were classified as used in the transportation of pigs between-farm. Using both reported and inferred classifications, we identified that 16\% of trucks transported feed, 24\% transported pigs to market, and 60\% transported pigs between farms. For trailers, 31\% transported feed, 23\% transported pigs to market, and 46\% transported pigs between farms. Based on the working roles, truck-and-trailer movement data were categorized and aggregated into three vehicle groups: pig, market, and feed.

\subsection*{Pig and vehicle movement network}

For the pig movement network, we constructed a directed network using the origin and destination of each daily movement recorded. The number of animals moved between farms was used as the edge weight. To allow networks to be benchmarked, this weight was normalized to a scale between 0 and 1 by dividing it by the maximum number of animals transported between farms during the study period (January 1, 2024, to December 30, 2025).

For vehicles, between-farm contact networks was generated by identifying vehicles' visits to farms, slaughterhouses, feed mills, and C\&D stations. Directed connections between farms were established based on the temporal sequence of vehicle visits, defined as a contact as a visit to one location followed by another. For each connection, we assigned a probability value ranging from 0 to 1 representing the likelihood of pathogen stability, based on ASF stability parameters on environmental surfaces. This probability was calculated using the elapsed time between visits and environmental temperature, which together modulate an exponential decay function describing pathogen stability over time \citep{mil-homens_assessment_2024, mazur-panasiuk_natural_2020, nuanualsuwan_persistence_2022}. Between-farm contacts were excluded if the vehicle underwent a successful C\&D process between visits; such contacts were defined as a vehicle visiting a C\&D station and remaining there for the required duration, under the assumption of an effective decontamination probability \citep{parker_evaluation_2025}. Additional details on visit identification, C\&D process, between-farm contact construction, and pathogen stability modeling are provided in \cite{galvis_role_2024}, \cite{galvis_mitigating_2025}, and \cite{galvis_descriptive_2026}. Using the generated farm-to-farm vehicle contacts, we constructed temporal directed vehicle movement networks in which edges represent time-ordered sequences of visits between locations, and edge weights correspond to the estimated ASF virus stability (selected for its current relevance to surveillance of potential introduction in the U.S.). From this framework, we generated overall vehicle networks for both trucks and trailers, with 24 and 9 million edges, respectively. In addition, networks were stratified according to vehicle roles previously described, resulting in three subsets: (i) pig transport between farms (truck or trailer pig network), (ii) pig transport to slaughterhouses (truck or trailer market network), and (iii) feed transport to farms (truck or trailer feed network).

\subsection*{Geodesic and road distance between farm networks}

To characterize spatial proximity between farms, we calculated the geodesic distance between all pairs of farms using their latitude and longitude coordinates. Geodesic distance represents the shortest distance over the Earth's surface between two geographic points \citep{hijmans_geosphere_2022}, and was estimated within each production company to avoid creating distance-based links between farms from different companies. After computing the pairwise distance matrix for each company, we retained only unique farm pairs and selected those within a predefined distance threshold of 10 km. This threshold was selected based on the risk distance on surveillance strategies for diseases such as ASF and foot-and-mouth disease in the U.S. \citep{usda_foot-and-mouth_2020, usda_african_2023}. Thus, we built a undirected geodesic distance network, in which links represent spatial proximity between farms without a specified direction, with 36634 edges. In addition, we estimated the road distance between farms. Road distances were calculated using the open source routing machine (OSRM) to identify the shortest travel route between two farm locations along the road network \citep{giraud_osrm_2022}. As in the geodesic distance analysis, road distances were computed only between farms within the same production company, and unique farm pairs were retained. Distances were expressed in kilometers and filtered according to a predefined 10 km threshold. As a result, we created a road undirected distance network with 15296 edges.

\subsection*{Network analysis}

We analyzed 11 networks as static representations: i) pig movements, ii) all truck movements, iii) feed truck movements, iv) pig truck movements, v) market truck movements, vi) all trailer movements, vii) feed trailer movements, viii) pig trailer movements, ix) market trailer movements, x) geodesic distance, and xi) road distance. In each network, multiple contacts, or edges, between the same pair of farms, or nodes, were aggregated by summing their weights. For each network, we quantified global connectivity by estimating the number of nodes, edges, and network density, defined as the proportion of observed connections to the total number of possible connections, ranging from 0 to 1 \citep{martinez-lopez_social_2009}. At the node level, we calculated centrality measures to characterize the role of individual farms within each network. Total degree was defined as the sum of all connections of a node, regardless of direction \citep{martinez-lopez_social_2009}. Strength was calculated as the sum of the weights of all edges connected to a node, representing the intensity of its connections \citep{barrat_architecture_2004}. Betweenness centrality was used to quantify node influence by measuring the proportion of shortest paths between all pairs of nodes that pass through a given node, thereby identifying farms that act as bridges within the network \citep{martinez-lopez_social_2009}. In addition, betweenness centrality was visualized on a logarithmic scale due to its highly skewed distribution. Differences between networks should therefore be interpreted in terms of orders of magnitude rather than absolute differences. Finally, we quantified the proportion of edges among farm production types for each network. For this, we aggregated and counted the edges according to the farms' origin and destination production types, and divided by the total number of edges in each network. We placed particular emphasis on edges directed toward sow farms because infections in these farms can have important economic and disease-transmission consequences \citep{cardenas_analyzing_2024, galvis_modeling_2022, galvis_modelling_2022, osemeke_economic_2025}. Accordingly, contacts directed to sow farms were classified as risky contacts.

\subsubsection*{Super-spreader analysis}

To identify farms with a high potential to contribute disproportionately to disease dissemination, we estimated a composite super-spreader score for each farm within each network. This score was calculated separately for directed and undirected networks to account for differences in the available centrality measures. For directed networks, which included pig, truck, and trailer movement networks, we used out-degree, out-strength, betweenness centrality, PageRank, and hub score. PageRank captures the relative importance of a farm based on the importance of its neighbors \citep{brin_anatomy_1998}, and hub score identifies farms that connect to highly influential destinations \citep{kleinberg_authoritative_1999}. For undirected distance-based networks, we used degree, strength, betweenness centrality, and eigenvector centrality, the latter representing a farm's influence based on the connectivity of the farms it is linked to \citep{bonacich_power_1987}. To make these measures comparable within each network, each metric was standardized as z-scores, a common approach for combining multiple network centrality measures into composite indices \citep{osgood_effects_2013, openshaw_map2k7_2020, puglisi_convergent_2025}. For a given metric $x$, the standardized value was calculated as

\begin{equation}
z_i = \frac{x_i - \bar{x}}{s_x}
\end{equation}

where $x_i$ is the value of the metric for farm $i$, $\bar{x}$ is the mean value of that metric across all farms in the network, and $s_x$ is the corresponding standard deviation.

For directed networks, the super-spreader score ($DS_i$) of farm $i$ was calculated as the mean of the five standardized centrality measures:

\begin{equation}
DS_i =
\frac{
z(\text{out\ degree}_i) +
z(\text{out\ strength}_i) +
z(\text{betweenness}_i) +
z(\text{pagerank}_i) +
z(\text{hub\ score}_i)
}{5}
\end{equation}

For undirected networks, the super-spreader score ($US_i$) was calculated as

\begin{equation}
US_i =
\frac{
z(\text{degree}_i) +
z(\text{strength}_i) +
z(\text{betweenness}_i) +
z(\text{eigenvector}_i)
}{4}
\end{equation}

After calculating the composite super-spreader score, farms were classified into three spreader-potential clusters using k-means clustering. The resulting clusters were ordered according to their mean composite score and labeled as low, medium, or high spreader potential. We then evaluated the distribution of farms classified as high spreader potential across production types. This classification was used only to facilitate visualization and comparison of farms across production types and networks. In addition, farms were ranked in descending order within each network, with farms with higher scores considered to have greater spreading potential. To evaluate the consistency of super-spreader identification across networks, we compared the farms ranked within the top 50 positions of each network, representing 1.4\% of the farms. This threshold was selected as a practical cutoff to capture a small subset of highly connected farms that could be prioritized for targeted surveillance and potentially serve as sentinels for early disease detection. However, this threshold is inherently arbitrary and does not correspond to a specific epidemiological cutoff. For each pair of networks, we calculated the number of farms shared by the two top-50 sets, the total number of unique farms across both sets, and the Jaccard similarity index. If $A$ and $B$ denote the sets of top-50 farms in two networks, the similarity metrics were calculated as

\begin{equation}
J(A,B) = \frac{|A \cap B|}{|A \cup B|}
\end{equation}

where $|A \cap B|$ is the number of farms common to both networks and $|A \cup B|$ is the number of farms present in at least one of the respective pair networks. All data processing and analyses were conducted using R version 4.2.3 and RStudio version 2024.4.2.764.1 \citep{r_core_team_r_2023, posit_team_rstudio_2024}.

\section*{Results}

\begin{figure}[H]
\centering
\includegraphics[width=0.99\textwidth]{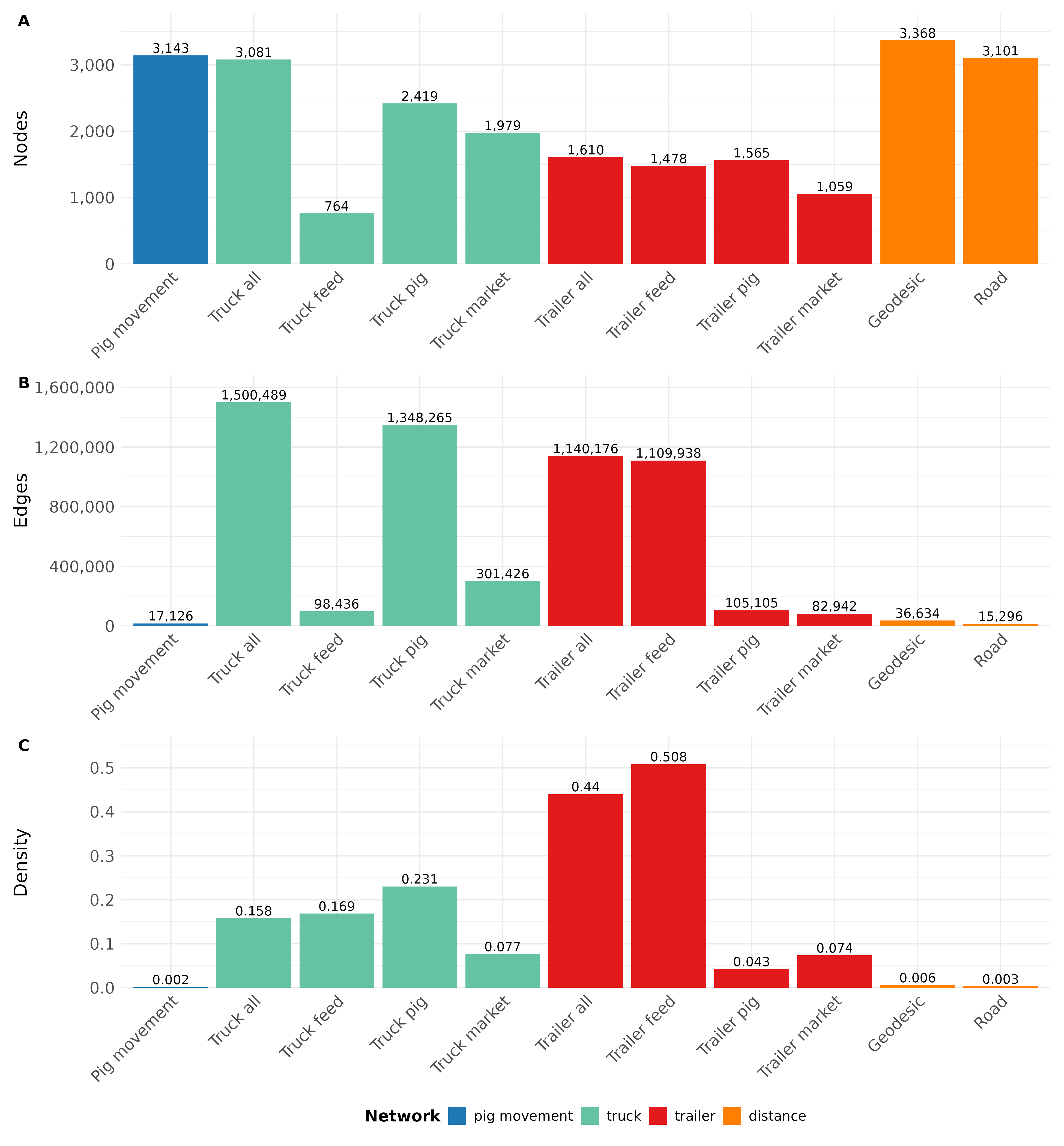}
\caption{Comparison of network size and connectivity across networks, including the A) number of nodes, B) number of edges, and C) network density.}
\label{fig:network_structure}
\end{figure}

Among the 11 between-farm contact networks, the geodesic distance network connected the highest number of nodes (Figure \ref{fig:network_structure}A). This was followed by pig movement (6.6\% fewer nodes), road distance (7.9\%), all truck movements (8.5\%), pig truck movements (28.2\%), market truck movements (41.2\%), all trailer movements (52.2\%), pig trailer movements (53.5\%), feed trailer movements (56.1\%), market trailer movements (68.6\%), and feed truck movements (77.3\% fewer nodes). This ranking differed when considering the number of edges (Figure \ref{fig:network_structure}B). The all truck movement network had the highest number of edges, followed by pig truck (10.1\% fewer edges), all trailer (24.0\%), feed trailer (26.0\%), market truck (79.9\%), pig trailer (93.0\%), feed truck (93.4\%), market trailer (94.5\%), geodesic distance (97.6\%), pig movement (98.9\%), and road distance networks (99.0\% fewer edges). In terms of network density (Figure \ref{fig:network_structure}C), the feed trailer movement network had the highest density, followed by all trailer and pig truck networks, both with 13.4\% lower density. This was followed by feed truck (66.8\% lower density), all trucks (68.9\%), market truck (84.9\%), market trailer (85.4\%), pig trailer (91.6\%), geodesic distance (98.7\%), road distance (99.4\%), and pig movement networks (99.7\% lower density).

\begin{figure}[H]
\centering
\includegraphics[width=0.99\textwidth]{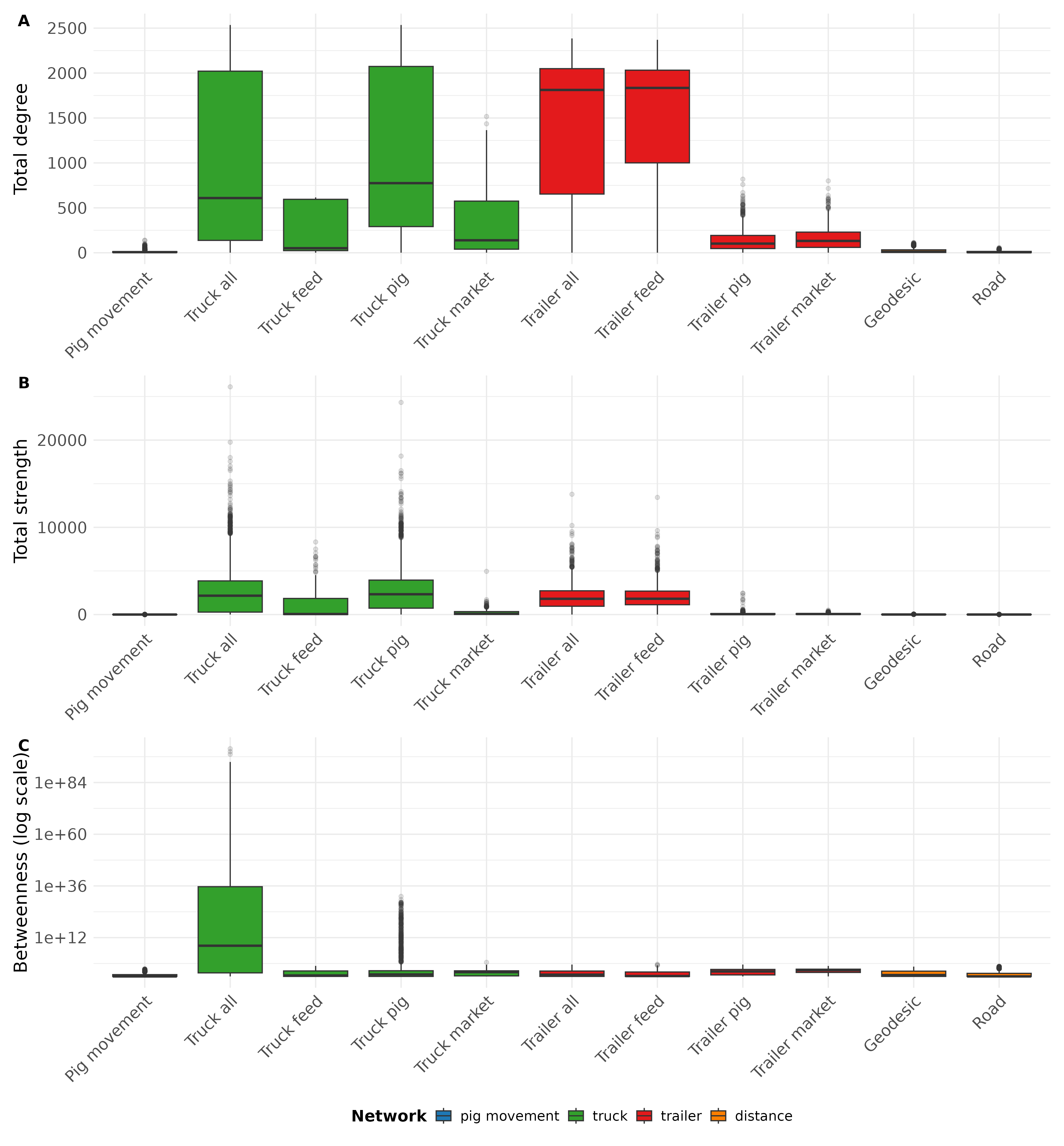}
\caption{Distribution of node-level centrality metrics across networks, including A) total degree, B) total strength, and C) betweenness centrality. Betweenness centrality is shown on a logarithmic scale to account for its highly skewed distribution.}
\label{fig:network_metrics}
\end{figure}

Relative to all trailer and feed-trailer movement networks, the median degree was 57.6\% to 66.8\% lower in the pig-truck and all-truck movement networks  (Figure \ref{fig:network_metrics}A). Larger reductions were observed in the market truck, market trailer, pig trailer, and feed truck movement networks, with reductions of 92.4\% to 97.2\%. The lowest median degree values were observed in the geodesic distance, pig movement, and road distance networks, which were 99.4\% to 99.7\% lower than the highest median degree values. 

For strength (Figure \ref{fig:network_metrics}B), pig truck and all truck movement networks showed the highest median values, followed by all trailer and feed trailer movement networks, which were 22.0\% lower. Market truck, feed truck, market trailer, and pig trailer movement networks had markedly lower median strength values, ranging from 97.4\% to 98.0\% lower. The geodesic distance, road distance, and pig movement networks showed the lowest median strength values, approximately 99.0\% lower than the highest median strength values. Regarding betweenness centrality (Figure \ref{fig:network_metrics}C), farms in the all truck movement network exhibited the highest median values, indicating a greater role in bridging connections between farms. This was followed, in descending order, by market trailer, pig trailer, market truck, pig truck, all trailer, geodesic distance, feed truck, feed trailer, pig movement, and road distance networks.

\begin{figure}[H]
\centering
\includegraphics[width=0.99\textwidth]{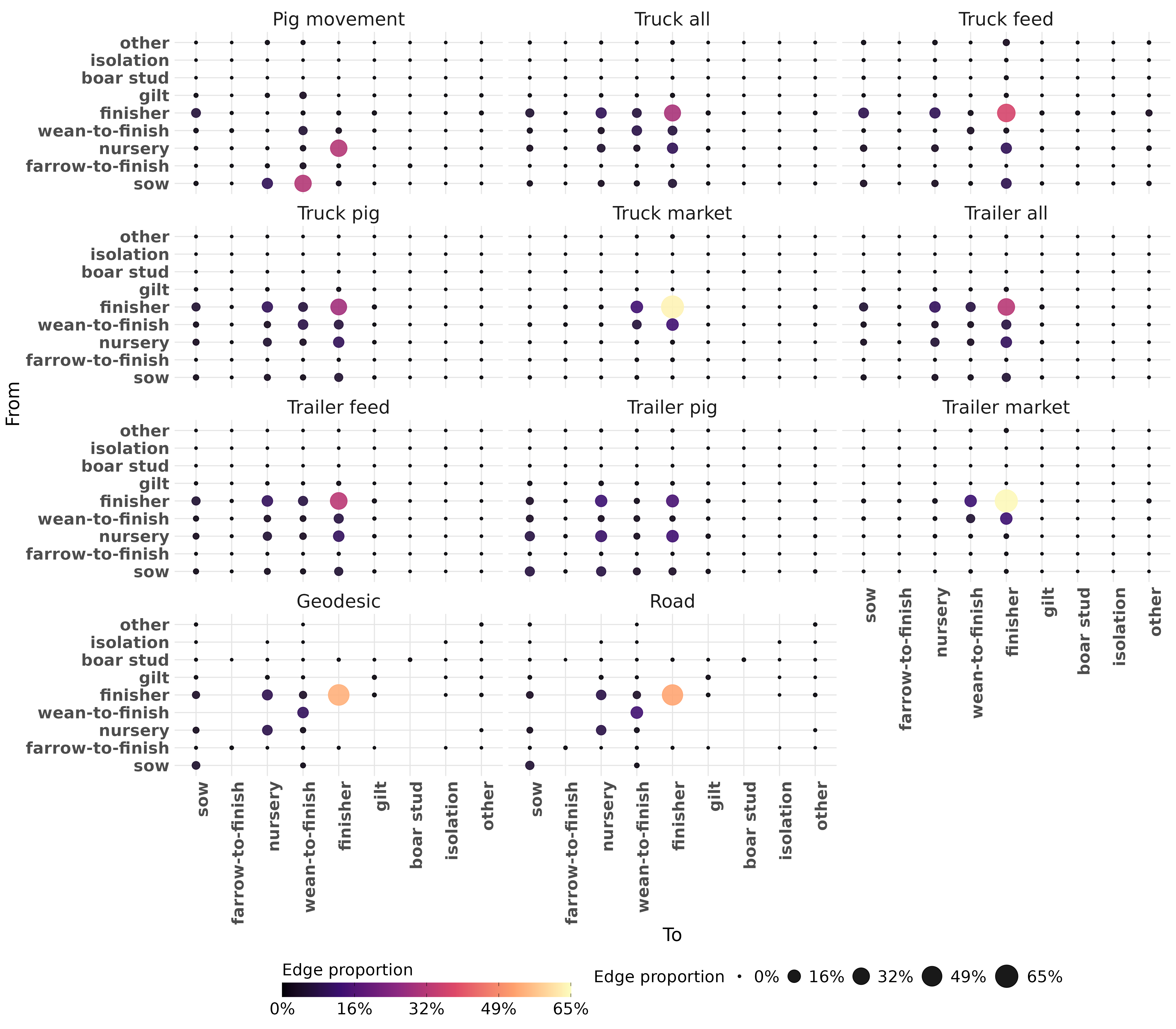}
\caption{Correlation plot showing the connectivity among swine farm production types across networks. Each panel represents a network, where circle size and color indicate the proportion of edges linking farm types. Proportions are calculated relative to the total number of edges within each network.}
\label{fig:cor_farmtype}
\end{figure}

Edges between finisher farms were the most frequent in vehicle and distance networks, with proportions ranging from 14\% in pig trailer movements to 64\% in market truck and trailer movement networks (Figure \ref{fig:cor_farmtype}). In contrast, in pig movement networks, edges from nursery to finisher farms were the most frequent, accounting for 32\% of all edges. Potentially risky connections, defined here as movements to sow farms, varied across networks. Finisher farms connected to sow farms with edge proportions ranging from 0.06\% in the market truck movement network to 8.7\% in the feed truck movement network (Figure \ref{fig:cor_farmtype}). In the pig movement network, edges from finisher to sow farms accounted for 6.2\%, while in the geodesic and road distance networks these proportions were 3.5\% and 2.8\%, respectively. Other production types also contributed to connections with sow farms. Nursery farms had edge proportions ranging from 0.02\% (market trailer) to 7.3\% (pig trailer), while sow-to-sow connections ranged from 0.01\% to 7.1\%, and wean-to-finish farms ranged from 0.15\% to 3.0\%, across networks. In contrast, gilt, other, farrow-to-finish, boar stud, and isolation farms showed relatively limited connectivity to sow farms, with maximum edge proportions between 0.04\% and 0.53\%, primarily observed in pig trailer and feed truck movement networks (Figure \ref{fig:cor_farmtype}).

\begin{figure}[H]
\centering
\includegraphics[width=0.99\textwidth,height=0.84\textheight,keepaspectratio]{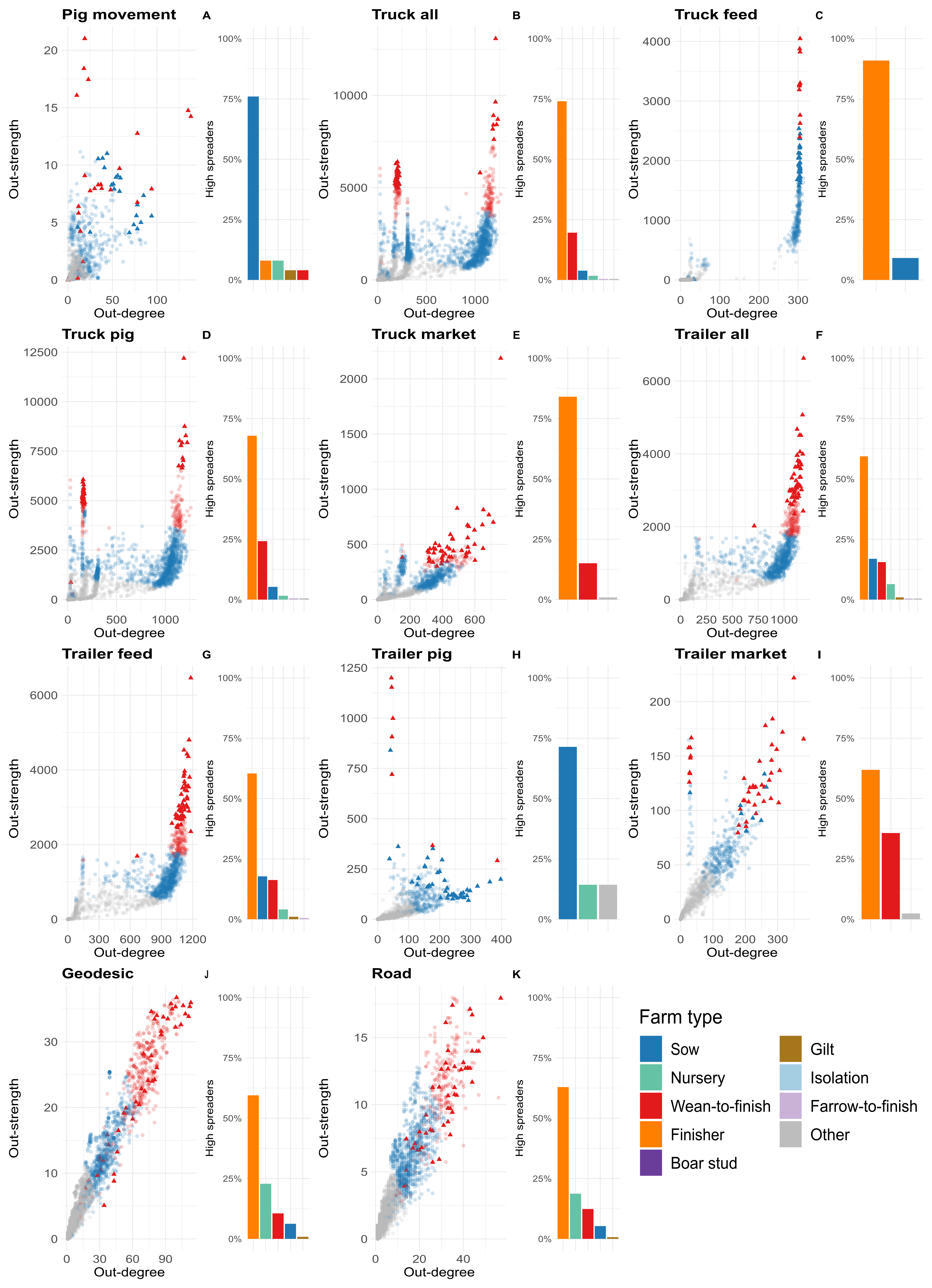}
\caption{Super-spreader nodes across networks. Each panel displays the relationship between node connectivity metrics, with out-degree on the x-axis and out-strength on the y-axis. Nodes are colored by spreader potential (low, medium, high), defined via k-means clustering on a composite score derived from multiple centrality measures. The 50 top-ranked nodes, based on this score, are highlighted with increased opacity and triangular symbols. The accompanying bar plot shows the proportion of nodes classified as high spreaders (color red) by production type within each network.}
\label{fig:super_spreader}
\end{figure}

In the identification of super-spreader farms using k-means clustering of the composite score into three groups (low, medium, and high), the geodesic distance network had the highest proportion of farms classified as high super-spreaders (12.8\%), followed by the road distance network (7.8\%) (Figure \ref{fig:super_spreader}). All truck, all trailer, feed trailer, pig truck, and market truck networks showed intermediate proportions of farms in the high super-spreader group, ranging from 3.1\% to 6.5\%. In contrast, market trailer, pig movement, feed truck, and pig trailer networks showed the lowest proportions, ranging from 0.2\% to 1.2\% of farms. Sow farms were the most common super-spreader farm type in the pig movement and pig trailer movement networks, representing 76\% and 71\% of farms classified as high super-spreaders, respectively (Figure \ref{fig:super_spreader}). In contrast, in the other vehicle type and distance networks, sow farms accounted for only 3.8\% to 17.8\% of high super-spreader farms. With the exception of the pig movement and pig trailer movement networks, finisher farms were the most common super-spreader farm type, representing between 59.5\% and 90.9\% of high super-spreader farms. The contribution of other farm types ranged from 0.4\% to 35.7\% of high super-spreader farms (Figure \ref{fig:super_spreader}).

\begin{figure}[H]
\centering
\includegraphics[width=0.99\textwidth]{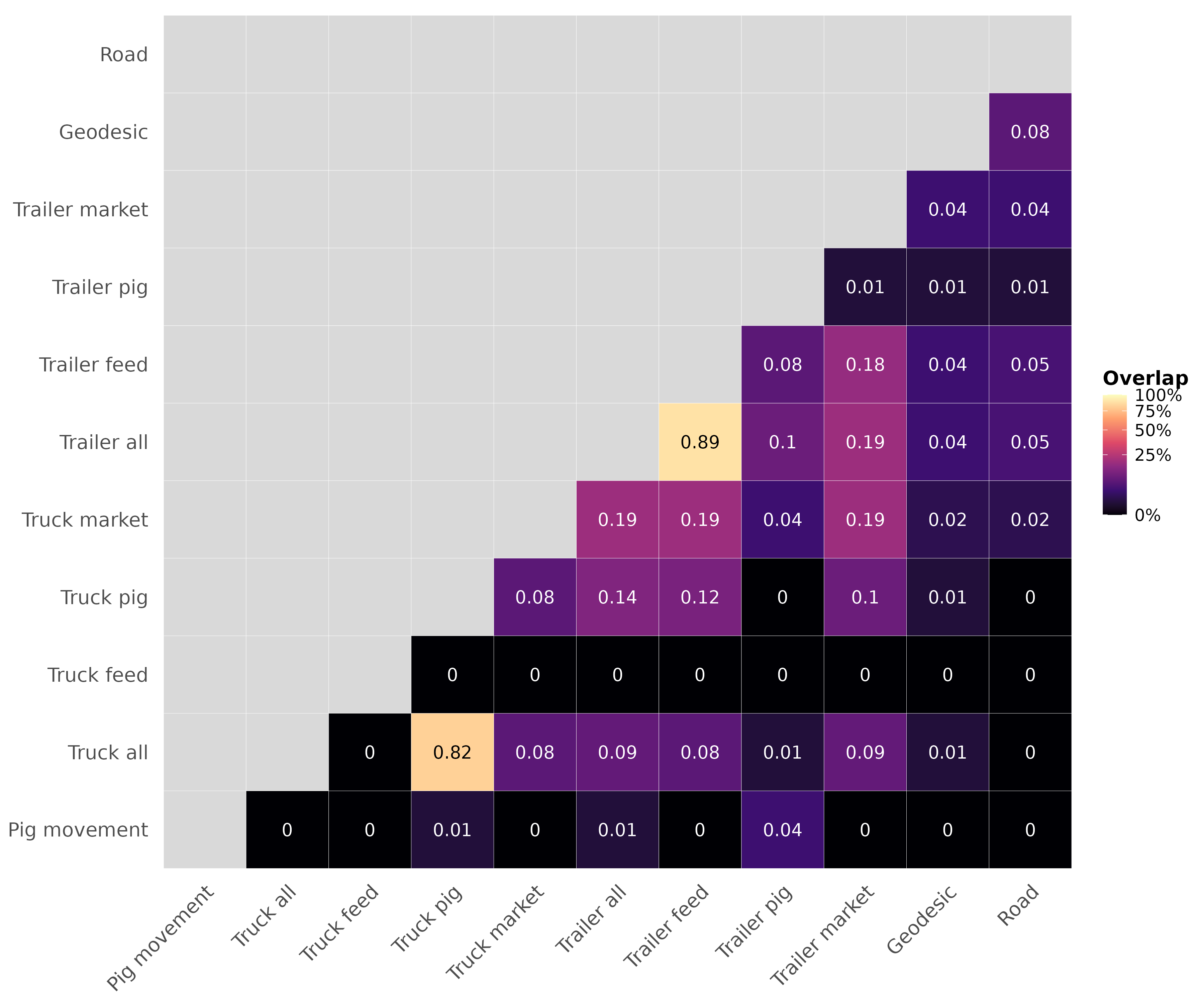}
\caption{Similarity of the 50 top-ranked nodes across networks. The heatmap shows pairwise similarity among the top 50 nodes in each network, with nodes ranked by a composite score derived from multiple centrality measures. Values represent the proportion of common nodes between each pair of networks, calculated using the Jaccard index.}
\label{fig:spreader_jaccard}
\end{figure}

When comparing the top 50 farms ranked by super-spreader potential, 89\% of these farms were shared between the all trailer and feed trailer movement networks, while 82\% were shared between the all truck and pig truck movement networks (Figure \ref{fig:spreader_jaccard}). In other comparisons of truck and trailer networks, the proportion of shared farms ranged from 0\% to 19\%. Pig movement networks showed limited similarity with other networks, with the proportion of shared farms ranging from 1\% to 4\% when compared with pig truck, all trailer, and pig trailer networks, and no shared farms in other comparisons (Figure \ref{fig:spreader_jaccard}). Interestingly, only 8\% of farms were shared between the geodesic and road distance networks (Figure \ref{fig:spreader_jaccard}). Moreover, these distance-based networks showed minimal similarity to the truck, trailer, and pig movement networks, with the proportion of shared farms ranging from 0\% to 4\%.

\section*{Discussion}

In this study, we evaluated 11 between-farm contact networks representing multiple transmission pathways and found substantial heterogeneity in their structure and connectivity patterns, with farms' risk of disease spread distributed heterogeneously. Vehicle-based networks, particularly the all-truck movement network, generated the highest connectivity, suggesting a major role in linking farms across large geographic areas, while distance-based and pig movement networks captured different aspects of farm interactions. Connectivity among production types also varied across networks, with finisher farms highly integrated with other farm types, particularly through vehicle and distance contact networks. In addition, sow farms, which play a central role in pig production \citep{cardenas_analyzing_2024, passafaro_network_2020}, were frequently reached across several networks, highlighting their vulnerability to pathogen introduction through multiple contact pathways. The identification of super-spreader farms was highly dependent on the network, with finisher farms dominating in most networks and sow farms emerging as key spreaders in pig-related movement networks. Importantly, the similarity of the 50 top-ranked super-spreader farms across networks was limited, indicating that high-risk farms are not consistently identified when, for instance, the truck network is not examined. Together, these findings demonstrate that disease transmission risk in swine systems is inherently multi-layered \citep{prezioso_network_2025, yi_multilayer_2022, kinsley_multilayer_2020}, and that reliance on a single network (e.g., pig movements) may lead to incomplete or biased surveillance strategies. Integrating multiple contact networks into surveillance frameworks could improve the identification of high-risk farms, enhance early detection through sentinel farms, and support targeted and effective disease control interventions.

Truck and trailer movement networks showed substantially higher connectivity and centrality role compared to pig movement and distance-based networks, reinforcing previous findings that indirect contacts by vehicle movements can strongly link swine farms and contribute to between-farm transmission \citep{galvis_role_2024, galvis_mitigating_2025, galvis_descriptive_2026, prezioso_network_2025, thakur_analysis_2016}. However, this does not imply that vehicle movements are the dominant transmission route \citep{galvis_modeling_2022, galvis_modelling_2022}. Direct pig movements involve transporting potentially infected animals between farms and are therefore likely to be among the most efficient pathways for pathogen dissemination \citep{passafaro_network_2020, cardenas_analyzing_2024, lentz_disease_2016}. Similarly, close spatial proximity between farms may facilitate indirect transmission through the movement of people, equipment, or vectors, further contributing to disease spread \citep{fleming_enhancing_2026, sanchez_spatiotemporal_2023}. The role of vehicle-mediated transmission depends on multiple factors, including pathogen stability in the environment, the probability of vehicle contamination, the likelihood of transferring contamination to farm environments, and the effectiveness and frequency of C\&D procedures \citep{parker_evaluation_2025, parker_evaluation_2025-1, mil-homens_assessment_2024, boniotti_porcine_2018, jacobs_stability_2010}. As a result, the overall probability of transmission through vehicles may be lower than that of direct animal movements or proximity-based mechanisms. Nevertheless, given the high connectivity observed in vehicle networks, a single contaminated vehicle, under favorable conditions, could expose many farms and increase the probability of disease introduction. Therefore, even if vehicle-mediated transmission occurs less frequently, its potential impact may be substantial \citep{neumann_is_2021}. Given the lack of infection-controlled field studies, our ability to further explain vehicle-based transmission is limited, and it becomes less of a priority for targeted interventions by stakeholders. 

It is widely known that infectious disease transmission in integrated commercial swine populations follows a predominantly vertical flow, from sow to downstream farms such as nurseries and finishers, likely driven by pig movements \citep{passafaro_network_2020, cardenas_analyzing_2024, lentz_disease_2016}. However, our results show that sow farms are widely connected to other production types via several networks, particularly through indirect contacts such as feed truck movements. Vehicle movements and proximity-based pathways highlight the potential for reverse direction on transmission or lateral transmission routes that are not captured by animal movements alone \citep{galvis_modeling_2022, galvis_modelling_2022}. Such contacts involving personnel, equipment, and vehicles may represent a significant risk of pathogen introduction into sow farms when biosecurity measures are insufficient \citep{fleming_enhancing_2026}. This is especially critical because infection in a sow farm often has substantial downstream consequences, given its central role in sourcing animals to the rest of the production system \citep{passafaro_network_2020, cardenas_analyzing_2024}. Once introduced, pathogens can propagate through nursery and finisher farms, with finisher farms potentially amplifying the spread further through their extensive connections across networks, as observed in our results. Although restricting pig movements to sow farms or increasing spatial separation is unrealistic in commercial production systems, reducing unnecessary indirect vehicle contacts could be a more feasible intervention \citep{galvis_mitigating_2025}. In particular, optimizing vehicle movement patterns and minimizing contact between production types, especially those from farms with lower biosecurity levels, could help reduce the risk of pathogen introduction into highly connected and vulnerable farms, such as sow farms.

Sow and finisher farms were most frequently identified as super-spreaders in our analysis. This finding is notable, as finisher farms are not typically considered primary super-spreaders; this role is more commonly attributed to sow and nursery farms due to the directional flow of pig movements and the limited reverse movement from finishers to other production types \citep{dorjee_network_2013, passafaro_network_2020, cardenas_analyzing_2024}. However, our results show that when truck and trailer movements and spatial proximity are taken into account, finisher farms play a major role in connecting farms within networks. Another important observation is that, although finisher farms were the most common farm type among super-spreaders, the sets of top-ranked farms were not the same across the 11 networks we analyzed (Figure \ref{fig:spreader_jaccard}). These findings support the assumption that integrating multiple contact networks is essential for identifying high-risk farms \citep{prezioso_network_2025, yi_multilayer_2022, kinsley_multilayer_2020}. At the same time, not all networks contribute equally to disease transmission \citep{galvis_modeling_2022, galvis_modelling_2022}, and enhancing surveillance strategies primarily on finisher farms may be insufficient given the variable transmission potential of vehicle and distance networks. A potential approach to integrating multiple network data is to assign transmission weights to each network based on its relative contribution to disease spread. For example, in ASFV, assigning greater weight to pig movement and distance-based networks may yield a more effective representation of transmission risk \citep{sykes_estimating_2023}. This strategy could support a framework for identifying super-spreader and sentinel farms across transmission pathways, thereby improving the design and effectiveness of disease surveillance programs.

\subsection*{Limitations and further remarks}

Despite the large volume of truck-and-trailer movement data included in this study, identifying vehicle contacts with farms and other facilities (e.g., C\&D stations) was limited by the availability of GPS tracking devices. As a result, vehicle movements without GPS data may not have been captured, potentially leading to an underestimation of connectivity within vehicle-based networks. In addition, the selection of the top 50 farms to define super-spreaders was based on a relatively small proportion of farms that could realistically be included in industry-based surveillance programs. This choice reflects practical constraints, as repeated monitoring, sampling, and diagnostic efforts are resource-intensive, and expanding surveillance to more farms may not be feasible \citep{galvis_estimating_2025}. Additional analysis is required to identify the optimal number of superspreader farms for disease surveillance; however, this was not conducted here, given the study's scope. Furthermore, this study was conducted using data from four swine companies with farms distributed across 18 U.S. states, and the analyses were performed globally across all networks. While this approach provides a comprehensive overview of connectivity patterns, it may mask important regional differences. Surveillance capacity, funding, and implementation strategies can vary substantially across states and swine production companies, suggesting that a combination of company- and state-level analyses could provide more targeted and operationally relevant insights for disease monitoring and control.

An additional consideration when interpreting our results is that vehicle movement networks were analyzed in an aggregated manner, combining all vehicles or subsets defined by vehicle type into a single network. In reality, contamination occurs at the level of individual vehicles, meaning that only a subset of movements may contribute to pathogen transmission \citep{parker_evaluation_2025, parker_evaluation_2025-1}. In contrast to pig movements, where aggregation is appropriate because each animal movement directly represents a potential transmission pathway within a defined production flow, vehicle-based transmission depends on stochastic contamination processes occurring at the level of individual vehicles. For example, a farm may become contaminated after a visit by one vehicle (Vehicle A), and a subsequent visit by another vehicle (Vehicle B) could, in theory, contaminate the second vehicle, thereby indirectly linking farms through sequential contacts by both vehicles. However, given the multiple conditions required for such events to occur (e.g., probability of a vehicle contaminating the farm, probability of acquiring contamination from the farm, pathogen stability, contamination of parking areas, C\&D effectiveness, and overall biosecurity), this sequence is likely to be rare. As a result, aggregating movements from different vehicles may overestimate potential transmission pathways by combining contacts that are not directly related in the transmission process. While this aggregation approach remains useful for capturing overall connectivity patterns, future work could benefit from constructing vehicle-specific contact networks, in which each vehicle defines its own temporal transmission pathway, providing a more realistic representation of transmission dynamics.

Our study compared 11 different networks for disease surveillance, made possible by the large volume of data collected from participating companies via RABapp\textsuperscript{\texttrademark} \citep{fleming_enhancing_2026}. These findings highlight the value of centralized data systems in facilitating the integration of multiple data sources and supporting more comprehensive disease surveillance efforts.

\section*{Conclusion}

This study demonstrates that between-farm connectivity varies markedly across transmission pathways, thus directly impacting the identification of high-risk farms and production types. The low similarity among farms classified as potential super-spreaders across pig movement, vehicle movement, and distance-based networks suggests that each network captures different aspects of disease transmission risk. Therefore, relying on a single contact structure may overlook important transmission routes. These findings emphasize the importance of integrating multiple contact networks to better inform disease surveillance and control strategies in swine and other animal production systems.

\section*{Acknowledgments}
The authors acknowledge the participating swine production companies for their collaboration.

\section*{Authors’ contributions}				
JAG and GM conceived the study. JAG designed the study. JAG collected vehicle movement and distance data. NCC processed animal movement data. JAG prepared the data and developed the analysis. JAG, NCC and GM wrote and edited the manuscript. All authors discussed the results and critically reviewed the manuscript. JAG and GM secured the funding.

\section*{Funding statement}
This study was funded by the Wean-to-Harvest Biosecurity Research Program, Swine Health Information Center (SHIC), and National Institute of Food and Agriculture-Data Science for Food and Agricultural Systems (DSFAS): Award Numbers:2024-67021-43841

\section*{Data Availability Statement}		
The data supporting this study's findings are not publicly available and are protected by confidential agreements; therefore, they are not available.

\bibliographystyle{elsarticle-harv}
\bibliography{network_comparison}

@misc{posit_team_rstudio_2024,
	address = {Boston, MA},
	title = {{RStudio}: {Integrated} {Development} {Environment} for {R}},
	url = {http://www.posit.co/},
	publisher = {Posit Software, PBC},
	author = {Posit team},
	year = {2024},
}

@misc{parker_evaluation_2025,
	title = {Evaluation of the infectivity of porcine epidemic diarrhea virus on swine vehicles after cleaning and disinfection},
	copyright = {Creative Commons Attribution Non Commercial Share Alike 4.0 International},
	url = {https://arxiv.org/abs/2512.06037},
	doi = {10.48550/ARXIV.2512.06037},
	urldate = {2025-12-10},
	publisher = {arXiv},
	author = {Parker, Taylor B. and Rahe, Michael C. and Meiklejohn, Kelly A. and Darrow, Bradford Sean and Galvis, Jason A. and Machado, Gustavo and Ferreira, Juliana Bonin},
	year = {2025},
	note = {Version Number: 1},
}

@misc{r_core_team_r_2023,
	address = {Vienna, Austria},
	title = {R: {A} {Language} and {Environment} for {Statistical} {Computing}},
	url = {https://www.R-project.org/},
	publisher = {R Foundation for Statistical Computing},
	author = {R Core Team},
	year = {2023},
}

@article{boniotti_porcine_2018,
	title = {Porcine {Epidemic} {Diarrhoea} {Virus} in {Italy}: {Disease} spread and the role of transportation},
	volume = {65},
	issn = {18651674},
	shorttitle = {Porcine {Epidemic} {Diarrhoea} {Virus} in {Italy}},
	url = {https://onlinelibrary.wiley.com/doi/10.1111/tbed.12974},
	doi = {10.1111/tbed.12974},
	language = {en},
	number = {6},
	urldate = {2022-05-05},
	journal = {Transboundary and Emerging Diseases},
	author = {Boniotti, Maria Beatrice and Papetti, Alice and Bertasio, Cristina and Giacomini, Enrico and Lazzaro, Massimiliano and Cerioli, Monica and Faccini, Silvia and Bonilauri, Paolo and Vezzoli, Fausto and Lavazza, Antonio and Alborali, Giovanni Loris},
	month = dec,
	year = {2018},
	pages = {1935--1942},
}

@article{abatzoglou_development_2013,
	title = {Development of gridded surface meteorological data for ecological applications and modelling},
	volume = {33},
	issn = {0899-8418, 1097-0088},
	url = {https://rmets.onlinelibrary.wiley.com/doi/10.1002/joc.3413},
	doi = {10.1002/joc.3413},
	language = {en},
	number = {1},
	urldate = {2025-08-08},
	journal = {International Journal of Climatology},
	author = {Abatzoglou, John T.},
	month = jan,
	year = {2013},
	pages = {121--131},
}

@article{parker_evaluation_2025-1,
	title = {Evaluation of porcine epidemic diarrhea virus {RNA} contamination on swine industry transportation vehicles},
	volume = {237},
	copyright = {https://www.elsevier.com/tdm/userlicense/1.0/},
	issn = {0167-5877},
	url = {https://linkinghub.elsevier.com/retrieve/pii/S0167587725000327},
	doi = {10.1016/j.prevetmed.2025.106447},
	language = {en},
	urldate = {2025-07-14},
	journal = {Preventive Veterinary Medicine},
	author = {Parker, Taylor B. and Meiklejohn, Kelly A. and Machado, Gustavo and Rahe, Michael and Darrow, Bradford Sean and Ferreira, Juliana Bonin},
	month = apr,
	year = {2025},
	note = {Publisher: Elsevier BV},
	pages = {106447},
}

@article{yoo_transmission_2021,
	title = {Transmission {Dynamics} of {African} {Swine} {Fever} {Virus}, {South} {Korea}, 2019},
	volume = {27},
	issn = {1080-6059},
	doi = {10.3201/eid2707.204230},
	language = {eng},
	number = {7},
	journal = {Emerging Infectious Diseases},
	author = {Yoo, Dae Sung and Kim, Younjung and Lee, Eune Sub and Lim, Jun Sik and Hong, Seong Keun and Lee, Il Seob and Jung, Chung Sik and Yoon, Ha Chung and Wee, Sung Hwan and Pfeiffer, Dirk U. and Fournié, Guillaume},
	month = jul,
	year = {2021},
	pmid = {34152953},
	pmcid = {PMC8237864},
	pages = {1909--1918},
}

@book{usda_african_2023,
	title = {African swine fever {Response} {Plan}: {The} {Red} {Book}},
	url = {https://www.aphis.usda.gov/animal_health/emergency_management/downloads/asf-responseplan.pdf},
	language = {en},
	author = {USDA},
	year = {2023},
}

@article{pujols_survivability_2014,
	title = {Survivability of porcine epidemic diarrhea virus ({PEDV}) in bovine plasma submitted to spray drying processing and held at different time by temperature storage conditions},
	volume = {174},
	issn = {03781135},
	url = {https://linkinghub.elsevier.com/retrieve/pii/S0378113514004994},
	doi = {10.1016/j.vetmic.2014.10.021},
	language = {en},
	number = {3-4},
	urldate = {2025-06-03},
	journal = {Veterinary Microbiology},
	author = {Pujols, Joan and Segalés, Joaquim},
	month = dec,
	year = {2014},
	pages = {427--432},
}

@article{quinonez-munoz_comparative_2024,
	title = {Comparative survival of five porcine reproductive and respiratory syndrome virus strains on six fomites},
	issn = {22310916, 09728988},
	url = {https://veterinaryworld.org/Vol.17/December-2024/8.php},
	doi = {10.14202/vetworld.2024.2774-2779},
	language = {en},
	urldate = {2025-06-03},
	journal = {Veterinary World},
	author = {Quinonez-Munoz, Angie and Sobhy, Nader M. and Goyal, Sagar M.},
	month = dec,
	year = {2024},
	pages = {2774--2779},
}

@article{mil-homens_assessment_2024,
	title = {Assessment of temperature and time on the survivability of porcine reproductive and respiratory syndrome virus ({PRRSV}) and porcine epidemic diarrhea virus ({PEDV}) on experimentally contaminated surfaces},
	volume = {19},
	issn = {1932-6203},
	url = {https://dx.plos.org/10.1371/journal.pone.0291181},
	doi = {10.1371/journal.pone.0291181},
	language = {en},
	number = {1},
	urldate = {2025-06-03},
	journal = {PLOS ONE},
	author = {Mil-Homens, Mafalda and Aljets, Ethan and Paiva, Rodrigo C. and Machado, Isadora and Cezar, Guilherme and Osemeke, Onyekachukwu and Moraes, Daniel and Jayaraman, Swaminathan and Brinning, Mckenna and Poeta Silva, Ana Paula and Tidgren, Lauren and Durflinger, Madison and Wilhelm, Mallory and Flores, Vivian and Frenier, Jolie and Linhares, Daniel and Zhang, Jianqiang and Holtkamp, Derald and Silva, Gustavo S.},
	editor = {Gladue, Douglas},
	month = jan,
	year = {2024},
	pages = {e0291181},
}

@article{jacobs_stability_2010,
	title = {Stability of \textit{{Porcine} {Reproductive} and {Respiratory} {Syndrome} virus} at {Ambient} {Temperatures}},
	volume = {22},
	copyright = {https://journals.sagepub.com/page/policies/text-and-data-mining-license},
	issn = {1040-6387, 1943-4936},
	url = {https://journals.sagepub.com/doi/10.1177/104063871002200216},
	doi = {10.1177/104063871002200216},
	language = {en},
	number = {2},
	urldate = {2025-06-03},
	journal = {Journal of Veterinary Diagnostic Investigation},
	author = {Jacobs, Anna C. and Hermann, Joseph R. and Muñoz-Zanzi, Claudia and Prickett, John R. and Roof, Michael B. and Yoon, Kyoung-Jin and Zimmerman, Jeffrey J.},
	month = mar,
	year = {2010},
	pages = {257--260},
}

@article{lentz_disease_2016,
	title = {Disease {Spread} through {Animal} {Movements}: {A} {Static} and {Temporal} {Network} {Analysis} of {Pig} {Trade} in {Germany}},
	volume = {11},
	issn = {1932-6203},
	shorttitle = {Disease {Spread} through {Animal} {Movements}},
	url = {https://dx.plos.org/10.1371/journal.pone.0155196},
	doi = {10.1371/journal.pone.0155196},
	language = {en},
	number = {5},
	urldate = {2026-04-20},
	journal = {PLOS ONE},
	author = {Lentz, Hartmut H. K. and Koher, Andreas and Hövel, Philipp and Gethmann, Jörn and Sauter-Louis, Carola and Selhorst, Thomas and Conraths, Franz J.},
	editor = {Boulinier, Thierry},
	month = may,
	year = {2016},
	pages = {e0155196},
}

@article{passafaro_network_2020,
	title = {Network analysis of swine movements in a multi-site pig production system in {Iowa}, {USA}},
	volume = {174},
	issn = {01675877},
	url = {https://linkinghub.elsevier.com/retrieve/pii/S0167587719303800},
	doi = {10.1016/j.prevetmed.2019.104856},
	language = {en},
	urldate = {2026-04-20},
	journal = {Preventive Veterinary Medicine},
	author = {Passafaro, Tiago L. and Fernandes, Arthur F.A. and Valente, Bruno D. and Williams, Noel H. and Rosa, Guilherme J.M.},
	month = jan,
	year = {2020},
	pages = {104856},
}

@article{acosta_network_2022,
	title = {Network analysis of pig movements in {Ecuador}: {Strengthening} surveillance of classical swine fever},
	volume = {69},
	issn = {1865-1674, 1865-1682},
	shorttitle = {Network analysis of pig movements in {Ecuador}},
	url = {https://onlinelibrary.wiley.com/doi/10.1111/tbed.14640},
	doi = {10.1111/tbed.14640},
	language = {en},
	number = {5},
	urldate = {2026-04-20},
	journal = {Transboundary and Emerging Diseases},
	author = {Acosta, Alfredo Javier and Cardenas, Nicolas Cespedes and Pisuna, Luis Miguel and Galvis, Jason A. and Vinueza, Rommel Lenin and Vasquez, Kleber Stalin and Grisi‐Filho, Jose Henrique and Amaku, Marcos and Gonçalves, Victor Salvador and Ferreira, Fernando},
	month = sep,
	year = {2022},
}

@article{machado_quantifying_2021,
	title = {Quantifying the dynamics of pig movements improves targeted disease surveillance and control plans},
	volume = {68},
	issn = {1865-1674, 1865-1682},
	url = {https://onlinelibrary.wiley.com/doi/10.1111/tbed.13841},
	doi = {10.1111/tbed.13841},
	language = {en},
	number = {3},
	urldate = {2026-04-20},
	journal = {Transboundary and Emerging Diseases},
	author = {Machado, Gustavo and Galvis, Jason Ardila and Lopes, Francisco Paulo Nunes and Voges, Joana and Medeiros, Antônio Augusto Rosa and Cárdenas, Nicolás Céspedes},
	month = may,
	year = {2021},
	pages = {1663--1675},
}

@article{cardenas_analyzing_2024,
	title = {Analyzing the intrastate and interstate swine movement network in the {United} {States}},
	volume = {230},
	issn = {01675877},
	url = {https://linkinghub.elsevier.com/retrieve/pii/S0167587724001508},
	doi = {10.1016/j.prevetmed.2024.106264},
	language = {en},
	urldate = {2026-04-20},
	journal = {Preventive Veterinary Medicine},
	author = {Cardenas, Nicolas C. and Valencio, Arthur and Sanchez, Felipe and O’Hara, Kathleen C. and Machado, Gustavo},
	month = sep,
	year = {2024},
	pages = {106264},
}

@article{galvis_modelling_2022,
	title = {Modelling and assessing additional transmission routes for porcine reproductive and respiratory syndrome virus: {Vehicle} movements and feed ingredients},
	volume = {69},
	issn = {1865-1674, 1865-1682},
	shorttitle = {Modelling and assessing additional transmission routes for porcine reproductive and respiratory syndrome virus},
	url = {https://onlinelibrary.wiley.com/doi/10.1111/tbed.14488},
	doi = {10.1111/tbed.14488},
	language = {en},
	number = {5},
	urldate = {2026-04-20},
	journal = {Transboundary and Emerging Diseases},
	author = {Galvis, Jason A. and Corzo, Cesar A. and Machado, Gustavo},
	month = sep,
	year = {2022},
}

@article{galvis_modeling_2022,
	title = {Modeling between-farm transmission dynamics of porcine epidemic diarrhea virus: {Characterizing} the dominant transmission routes},
	volume = {208},
	issn = {01675877},
	shorttitle = {Modeling between-farm transmission dynamics of porcine epidemic diarrhea virus},
	url = {https://linkinghub.elsevier.com/retrieve/pii/S0167587722001921},
	doi = {10.1016/j.prevetmed.2022.105759},
	language = {en},
	urldate = {2026-04-20},
	journal = {Preventive Veterinary Medicine},
	author = {Galvis, Jason A. and Corzo, Cesar A. and Prada, Joaquín M. and Machado, Gustavo},
	month = nov,
	year = {2022},
	pages = {105759},
}

@article{olesen_transmission_2017,
	title = {Transmission of {African} swine fever virus from infected pigs by direct contact and aerosol routes},
	volume = {211},
	issn = {03781135},
	url = {https://linkinghub.elsevier.com/retrieve/pii/S0378113517307757},
	doi = {10.1016/j.vetmic.2017.10.004},
	language = {en},
	urldate = {2026-04-20},
	journal = {Veterinary Microbiology},
	author = {Olesen, Ann Sofie and Lohse, Louise and Boklund, Anette and Halasa, Tariq and Gallardo, Carmina and Pejsak, Zygmunt and Belsham, Graham J. and Rasmussen, Thomas Bruun and Bøtner, Anette},
	month = nov,
	year = {2017},
	pages = {92--102},
}

@article{guinat_transmission_2016,
	title = {Transmission routes of {African} swine fever virus to domestic pigs: current knowledge and future research directions},
	volume = {178},
	copyright = {http://creativecommons.org/licenses/by/4.0/},
	issn = {0042-4900, 2042-7670},
	shorttitle = {Transmission routes of {African} swine fever virus to domestic pigs},
	url = {https://bvajournals.onlinelibrary.wiley.com/doi/10.1136/vr.103593},
	doi = {10.1136/vr.103593},
	language = {en},
	number = {11},
	urldate = {2026-04-20},
	journal = {Veterinary Record},
	author = {Guinat, Claire and Gogin, Andrey and Blome, Sandra and Keil, Guenther and Pollin, Reiko and Pfeiffer, Dirk U. and Dixon, Linda},
	month = mar,
	year = {2016},
	pages = {262--267},
}

@article{gebhardt_sampling_2022,
	title = {Sampling and detection of {African} swine fever virus within a feed manufacturing and swine production system},
	volume = {69},
	issn = {1865-1674, 1865-1682},
	url = {https://onlinelibrary.wiley.com/doi/10.1111/tbed.14335},
	doi = {10.1111/tbed.14335},
	language = {en},
	number = {1},
	urldate = {2026-04-20},
	journal = {Transboundary and Emerging Diseases},
	author = {Gebhardt, Jordan T. and Dritz, Steve S. and Elijah, C. Grace and Jones, Cassandra K. and Paulk, Chad B. and Woodworth, Jason C.},
	month = jan,
	year = {2022},
	pages = {103--114},
}

@article{vanderwaal_role_2018,
	title = {Role of animal movement and indirect contact among farms in transmission of porcine epidemic diarrhea virus},
	volume = {24},
	issn = {17554365},
	url = {https://linkinghub.elsevier.com/retrieve/pii/S1755436517301755},
	doi = {10.1016/j.epidem.2018.04.001},
	language = {en},
	urldate = {2026-04-20},
	journal = {Epidemics},
	author = {VanderWaal, Kimberly and Perez, Andres and Torremorrell, Montse and Morrison, Robert M. and Craft, Meggan},
	month = sep,
	year = {2018},
	pages = {67--75},
}

@article{galvis_role_2024,
	title = {The role of vehicle movement in swine disease dissemination: {Novel} method accounting for pathogen stability and vehicle cleaning effectiveness uncertainties},
	volume = {226},
	issn = {01675877},
	shorttitle = {The role of vehicle movement in swine disease dissemination},
	url = {https://linkinghub.elsevier.com/retrieve/pii/S0167587724000540},
	doi = {10.1016/j.prevetmed.2024.106168},
	language = {en},
	urldate = {2026-04-20},
	journal = {Preventive Veterinary Medicine},
	author = {Galvis, Jason A. and Machado, Gustavo},
	month = may,
	year = {2024},
	pages = {106168},
}

@article{galvis_mitigating_2025,
	title = {Mitigating between-farm disease transmission through simulating vehicle rerouting and enhanced cleaning and disinfection protocols},
	volume = {244},
	issn = {01675877},
	url = {https://linkinghub.elsevier.com/retrieve/pii/S0167587725002351},
	doi = {10.1016/j.prevetmed.2025.106650},
	language = {en},
	urldate = {2026-04-20},
	journal = {Preventive Veterinary Medicine},
	author = {Galvis, Jason A. and Corzo, Cesar A. and Machado, Gustavo},
	month = nov,
	year = {2025},
	pages = {106650},
}

@article{sykes_estimating_2023,
	title = {Estimating the effectiveness of control actions on {African} swine fever transmission in commercial swine populations in the {United} {States}},
	volume = {217},
	issn = {01675877},
	url = {https://linkinghub.elsevier.com/retrieve/pii/S0167587723001265},
	doi = {10.1016/j.prevetmed.2023.105962},
	language = {en},
	urldate = {2026-04-20},
	journal = {Preventive Veterinary Medicine},
	author = {Sykes, Abagael L. and Galvis, Jason A. and O’Hara, Kathleen C. and Corzo, Cesar and Machado, Gustavo},
	month = aug,
	year = {2023},
	pages = {105962},
}

@article{sanchez_spatiotemporal_2023,
	title = {Spatiotemporal relative risk distribution of porcine reproductive and respiratory syndrome virus in the {United} {States}},
	volume = {10},
	issn = {2297-1769},
	url = {https://www.frontiersin.org/articles/10.3389/fvets.2023.1158306/full},
	doi = {10.3389/fvets.2023.1158306},
	urldate = {2026-04-20},
	journal = {Frontiers in Veterinary Science},
	author = {Sanchez, Felipe and Galvis, Jason A. and Cardenas, Nicolas C. and Corzo, Cesar and Jones, Christopher and Machado, Gustavo},
	month = jun,
	year = {2023},
	pages = {1158306},
}

@article{fleming_enhancing_2026,
	title = {Enhancing {U}.{S}. swine farm preparedness for infectious foreign animal diseases with rapid access to biosecurity information},
	volume = {248},
	issn = {01675877},
	url = {https://linkinghub.elsevier.com/retrieve/pii/S0167587725003502},
	doi = {10.1016/j.prevetmed.2025.106765},
	language = {en},
	urldate = {2026-04-20},
	journal = {Preventive Veterinary Medicine},
	author = {Fleming, Christian and Mills, Kelsey and Cardenas, Nicolas C. and Galvis, Jason A. and Corzo, Cesar and Ebling, Denílson Dos Santos and Machado, Gustavo},
	month = mar,
	year = {2026},
	pages = {106765},
}

@article{alarcon_biosecurity_2021,
	title = {Biosecurity in pig farms: a review},
	volume = {7},
	issn = {2055-5660},
	shorttitle = {Biosecurity in pig farms},
	url = {https://porcinehealthmanagement.biomedcentral.com/articles/10.1186/s40813-020-00181-z},
	doi = {10.1186/s40813-020-00181-z},
	language = {en},
	number = {1},
	urldate = {2026-04-21},
	journal = {Porcine Health Management},
	author = {Alarcón, Laura Valeria and Allepuz, Alberto and Mateu, Enric},
	month = jan,
	year = {2021},
	pages = {5},
}

@misc{hijmans_geosphere_2022,
	title = {geosphere: {Spherical} {Trigonometry}},
	url = {https://CRAN.R-project.org/package=geosphere},
	author = {Hijmans, Robert J.},
	year = {2022},
}

@article{thakur_analysis_2016,
	title = {Analysis of {Swine} {Movement} in {Four} {Canadian} {Regions}: {Network} {Structure} and {Implications} for {Disease} {Spread}},
	volume = {63},
	copyright = {http://doi.wiley.com/10.1002/tdm\_license\_1.1},
	issn = {18651674},
	shorttitle = {Analysis of {Swine} {Movement} in {Four} {Canadian} {Regions}},
	url = {https://onlinelibrary.wiley.com/doi/10.1111/tbed.12225},
	doi = {10.1111/tbed.12225},
	language = {en},
	number = {1},
	urldate = {2026-04-21},
	journal = {Transboundary and Emerging Diseases},
	author = {Thakur, K. K. and Revie, C. W. and Hurnik, D. and Poljak, Z. and Sanchez, J.},
	month = feb,
	year = {2016},
	pages = {e14--e26},
}

@article{andraud_modelling_2022,
	title = {Modelling {African} swine fever virus spread in pigs using time‐respective network data: {Scientific} support for decision makers},
	volume = {69},
	issn = {1865-1674, 1865-1682},
	shorttitle = {Modelling {African} swine fever virus spread in pigs using time‐respective network data},
	url = {https://onlinelibrary.wiley.com/doi/10.1111/tbed.14550},
	doi = {10.1111/tbed.14550},
	language = {en},
	number = {5},
	urldate = {2026-04-21},
	journal = {Transboundary and Emerging Diseases},
	author = {Andraud, Mathieu and Hammami, Pachka and H. Hayes, Brandon and A. Galvis, Jason and Vergne, Timothée and Machado, Gustavo and Rose, Nicolas},
	month = sep,
	year = {2022},
}

@article{masserdotti_role_2026,
	title = {The {Role} of {Short} {Journey} {Transportation} in the {Spreading} of {Swine} {Pathogens} and {Antimicrobial}‐{Resistant} {Bacteria}},
	volume = {2026},
	issn = {1865-1674, 1865-1682},
	url = {https://onlinelibrary.wiley.com/doi/10.1155/tbed/5600771},
	doi = {10.1155/tbed/5600771},
	language = {en},
	number = {1},
	urldate = {2026-04-21},
	journal = {Transboundary and Emerging Diseases},
	author = {Masserdotti, Marta and Formenti, Nicoletta and Donneschi, Anna and Guarneri, Flavia and Scali, Federico and Romeo, Claudia and Giacomini, Enrico and Bertasio, Cristina and Boniotti, Maria Beatrice and Alborali, Giovanni Loris and Luzzago, Camilla},
	editor = {Saini, Esha},
	month = jan,
	year = {2026},
	pages = {5600771},
}

@article{prezioso_network_2025,
	title = {A network evaluation of human and animal movement data across multiple swine farm systems in {North} {America}},
	volume = {234},
	issn = {01675877},
	url = {https://linkinghub.elsevier.com/retrieve/pii/S0167587724002563},
	doi = {10.1016/j.prevetmed.2024.106370},
	language = {en},
	urldate = {2026-04-21},
	journal = {Preventive Veterinary Medicine},
	author = {Prezioso, Tara and Boakes, Alicia and Wrathall, Jeff and Reger, W. Jonas and Bhowmick, Suman and Smith, Rebecca Lee},
	month = jan,
	year = {2025},
	pages = {106370},
}

@misc{galvis_descriptive_2026,
	title = {Descriptive and risk analysis of vehicle movements linked to porcine reproductive and respiratory syndrome and porcine epidemic diarrhea transmission in {US} commercial swine farms},
	copyright = {Creative Commons Attribution Non Commercial Share Alike 4.0 International},
	url = {https://arxiv.org/abs/2601.18819},
	doi = {10.48550/ARXIV.2601.18819},
	urldate = {2026-04-22},
	publisher = {arXiv},
	author = {Galvis, Jason A. and Parker, Taylor B. and Corzo, Cesar A. and Ferreira, Juliana B. and Meiklejohn, Kelly A. and Machado, Gustavo},
	year = {2026},
	note = {Version Number: 1},
}

@article{giraud_osrm_2022,
	title = {osrm: {Interface} {Between} {R} and the {OpenStreetMap}-{BasedRouting} {Service} {OSRM}},
	volume = {7},
	copyright = {http://creativecommons.org/licenses/by/4.0/},
	issn = {2475-9066},
	shorttitle = {osrm},
	url = {https://joss.theoj.org/papers/10.21105/joss.04574},
	doi = {10.21105/joss.04574},
	number = {78},
	urldate = {2026-04-22},
	journal = {Journal of Open Source Software},
	author = {Giraud, Timothée},
	month = oct,
	year = {2022},
	pages = {4574},
}

@article{martinez-lopez_social_2009,
	title = {Social {Network} {Analysis}. {Review} of {General} {Concepts} and {Use} in {Preventive} {Veterinary} {Medicine}},
	volume = {56},
	copyright = {http://doi.wiley.com/10.1002/tdm\_license\_1.1},
	issn = {18651674, 18651682},
	url = {https://onlinelibrary.wiley.com/doi/10.1111/j.1865-1682.2009.01073.x},
	doi = {10.1111/j.1865-1682.2009.01073.x},
	language = {en},
	number = {4},
	urldate = {2026-04-22},
	journal = {Transboundary and Emerging Diseases},
	author = {Martínez-López, B. and Perez, A. M. and Sánchez-Vizcaíno, J. M.},
	month = may,
	year = {2009},
	pages = {109--120},
}

@article{barrat_architecture_2004,
	title = {The architecture of complex weighted networks},
	volume = {101},
	issn = {0027-8424, 1091-6490},
	url = {https://pnas.org/doi/full/10.1073/pnas.0400087101},
	doi = {10.1073/pnas.0400087101},
	language = {en},
	number = {11},
	urldate = {2026-04-22},
	journal = {Proceedings of the National Academy of Sciences},
	author = {Barrat, A. and Barthélemy, M. and Pastor-Satorras, R. and Vespignani, A.},
	month = mar,
	year = {2004},
	pages = {3747--3752},
}

@article{brin_anatomy_1998,
	title = {The anatomy of a large-scale hypertextual {Web} search engine},
	volume = {30},
	copyright = {https://www.elsevier.com/tdm/userlicense/1.0/},
	issn = {01697552},
	url = {https://linkinghub.elsevier.com/retrieve/pii/S016975529800110X},
	doi = {10.1016/S0169-7552(98)00110-X},
	language = {en},
	number = {1-7},
	urldate = {2026-04-22},
	journal = {Computer Networks and ISDN Systems},
	author = {Brin, Sergey and Page, Lawrence},
	month = apr,
	year = {1998},
	pages = {107--117},
}

@article{kleinberg_authoritative_1999,
	title = {Authoritative sources in a hyperlinked environment},
	volume = {46},
	issn = {0004-5411, 1557-735X},
	url = {https://dl.acm.org/doi/10.1145/324133.324140},
	doi = {10.1145/324133.324140},
	language = {en},
	number = {5},
	urldate = {2026-04-22},
	journal = {Journal of the ACM},
	author = {Kleinberg, Jon M.},
	month = sep,
	year = {1999},
	pages = {604--632},
}

@article{bonacich_power_1987,
	title = {Power and {Centrality}: {A} {Family} of {Measures}},
	volume = {92},
	issn = {0002-9602, 1537-5390},
	shorttitle = {Power and {Centrality}},
	url = {https://www.journals.uchicago.edu/doi/10.1086/228631},
	doi = {10.1086/228631},
	language = {en},
	number = {5},
	urldate = {2026-04-22},
	journal = {American Journal of Sociology},
	author = {Bonacich, Phillip},
	month = mar,
	year = {1987},
	pages = {1170--1182},
}

@article{galvis_estimating_2025,
	title = {Estimating sampling and laboratory capacity for a simulated {African} swine fever outbreak in the {United} {States}},
	volume = {239},
	issn = {01675877},
	url = {https://linkinghub.elsevier.com/retrieve/pii/S016758772500114X},
	doi = {10.1016/j.prevetmed.2025.106529},
	language = {en},
	urldate = {2026-04-23},
	journal = {Preventive Veterinary Medicine},
	author = {Galvis, Jason A. and Satici, Muhammed Y. and Sykes, Abagael L. and O’Hara, Kathleen C. and Rochette, Lisa and Roberts, David and Machado, Gustavo},
	month = jun,
	year = {2025},
	pages = {106529},
}

@article{neumann_is_2021,
	title = {Is transportation a risk factor for {African} swine fever transmission in {Australia}: a review},
	volume = {99},
	issn = {0005-0423, 1751-0813},
	shorttitle = {Is transportation a risk factor for {African} swine fever transmission in {Australia}},
	url = {https://onlinelibrary.wiley.com/doi/10.1111/avj.13106},
	doi = {10.1111/avj.13106},
	language = {en},
	number = {11},
	urldate = {2026-04-23},
	journal = {Australian Veterinary Journal},
	author = {Neumann, Ej and Hall, Wf and Dahl, J and Hamilton, D and Kurian, A},
	month = nov,
	year = {2021},
	pages = {459--468},
}

@article{osgood_effects_2013,
	title = {Effects of {PROSPER} on the {Influence} {Potential} of {Prosocial} {Versus} {Antisocial} {Youth} in {Adolescent} {Friendship} {Networks}},
	volume = {53},
	copyright = {https://www.elsevier.com/tdm/userlicense/1.0/},
	issn = {1054139X},
	url = {https://linkinghub.elsevier.com/retrieve/pii/S1054139X13001080},
	doi = {10.1016/j.jadohealth.2013.02.013},
	language = {en},
	number = {2},
	urldate = {2026-04-23},
	journal = {Journal of Adolescent Health},
	author = {Osgood, D. Wayne and Feinberg, Mark E. and Gest, Scott D. and Moody, James and Ragan, Daniel T. and Spoth, Richard and Greenberg, Mark and Redmond, Cleve},
	month = aug,
	year = {2013},
	pages = {174--179},
}

@article{openshaw_map2k7_2020,
	title = {Map2k7 {Haploinsufficiency} {Induces} {Brain} {Imaging} {Endophenotypes} and {Behavioral} {Phenotypes} {Relevant} to {Schizophrenia}},
	volume = {46},
	issn = {1745-1701},
	doi = {10.1093/schbul/sbz044},
	language = {eng},
	number = {1},
	journal = {Schizophrenia Bulletin},
	author = {Openshaw, Rebecca L. and Thomson, David M. and Thompson, Rhiannon and Penninger, Josef M. and Pratt, Judith A. and Morris, Brian J. and Dawson, Neil},
	month = jan,
	year = {2020},
	pmid = {31219577},
	pmcid = {PMC6942167},
	pages = {211--223},
}

@article{puglisi_convergent_2025,
	title = {Convergent causal mapping unravels distinct frontal networks for visuospatial selective attention},
	volume = {17},
	issn = {2041-1723},
	url = {https://www.nature.com/articles/s41467-025-67381-5},
	doi = {10.1038/s41467-025-67381-5},
	language = {en},
	number = {1},
	urldate = {2026-04-23},
	journal = {Nature Communications},
	author = {Puglisi, Guglielmo and Viganò, Luca and Leonetti, Antonella and Rossi, Marco and Sciortino, Tommaso and Conti Nibali, Marco and Gay, Lorenzo Gabriel and Mollica, Luca and Fornia, Luca and Cerri, Gabriella and Bello, Lorenzo},
	month = dec,
	year = {2025},
	pages = {659},
}

@article{kinsley_multilayer_2020,
	title = {Multilayer and {Multiplex} {Networks}: {An} {Introduction} to {Their} {Use} in {Veterinary} {Epidemiology}},
	volume = {7},
	issn = {2297-1769},
	shorttitle = {Multilayer and {Multiplex} {Networks}},
	url = {https://www.frontiersin.org/article/10.3389/fvets.2020.00596/full},
	doi = {10.3389/fvets.2020.00596},
	urldate = {2026-04-24},
	journal = {Frontiers in Veterinary Science},
	author = {Kinsley, Amy C. and Rossi, Gianluigi and Silk, Matthew J. and VanderWaal, Kimberly},
	month = sep,
	year = {2020},
	pages = {596},
}

@article{yi_multilayer_2022,
	title = {Multilayer network analysis of {FMD} transmission and containment among beef cattle farms},
	volume = {12},
	issn = {2045-2322},
	url = {https://www.nature.com/articles/s41598-022-19981-0},
	doi = {10.1038/s41598-022-19981-0},
	language = {en},
	number = {1},
	urldate = {2026-04-24},
	journal = {Scientific Reports},
	author = {Yi, Chunlin and Yang, Qihui and Scoglio, Caterina M.},
	month = sep,
	year = {2022},
	pages = {15679},
}

@article{dee_experimental_2004,
	title = {An experimental model to evaluate the role of transport vehicles as a source of transmission of porcine reproductive and respiratory syndrome virus to susceptible pigs},
	volume = {68},
	issn = {0830-9000},
	language = {eng},
	number = {2},
	journal = {Canadian Journal of Veterinary Research = Revue Canadienne De Recherche Veterinaire},
	author = {Dee, Scott A. and Deen, John and Otake, Satoshi and Pijoan, Carlos},
	month = apr,
	year = {2004},
	pages = {128--133},
}

@article{mannion_investigation_2008,
	title = {An {Investigation} into the {Efficacy} of {Washing} {Trucks} {Following} the {Transportation} of {Pigs}—{A} \textit{{Salmonella}} {Perspective}},
	volume = {5},
	copyright = {https://journals.sagepub.com/page/policies/text-and-data-mining-license},
	issn = {1535-3141, 1556-7125},
	url = {https://journals.sagepub.com/doi/10.1089/fpd.2007.0069},
	doi = {10.1089/fpd.2007.0069},
	language = {en},
	number = {3},
	urldate = {2026-04-27},
	journal = {Foodborne Pathogens and Disease},
	author = {Mannion, Celine and Egan, John and Lynch, Brendan P. and Fanning, Seamus and Leonard, Nola},
	month = jun,
	year = {2008},
	pages = {261--271},
}

@article{houston_evaluation_2024,
	title = {Evaluation of {Truck} {Cab} {Decontamination} {Procedures} following {Inoculation} with {Porcine} {Epidemic} {Diarrhea} {Virus} and {Porcine} {Reproductive} and {Respiratory} {Syndrome} {Virus}},
	volume = {14},
	issn = {2076-2615},
	url = {https://www.mdpi.com/2076-2615/14/2/280},
	doi = {10.3390/ani14020280},
	language = {en},
	number = {2},
	urldate = {2026-04-27},
	journal = {Animals},
	author = {Houston, Grace E. and Jones, Cassandra K. and Evans, Caitlin and Otott, Haley K. and Stark, Charles R. and Bai, Jianfa and Poulsen Porter, Elizabeth G. and De Almeida, Marcelo N. and Zhang, Jianqiang and Gauger, Phillip C. and Blomme, Allison K. and Woodworth, Jason C. and Paulk, Chad B. and Gebhardt, Jordan T.},
	month = jan,
	year = {2024},
	pages = {280},
}

@article{mazur-panasiuk_natural_2020,
	title = {Natural inactivation of {African} swine fever virus in tissues: {Influence} of temperature and environmental conditions on virus survival},
	volume = {242},
	issn = {03781135},
	shorttitle = {Natural inactivation of {African} swine fever virus in tissues},
	url = {https://linkinghub.elsevier.com/retrieve/pii/S0378113519313240},
	doi = {10.1016/j.vetmic.2020.108609},
	language = {en},
	urldate = {2026-04-27},
	journal = {Veterinary Microbiology},
	author = {Mazur-Panasiuk, Natalia and Woźniakowski, Grzegorz},
	month = mar,
	year = {2020},
	pages = {108609},
}

@article{nuanualsuwan_persistence_2022,
	title = {Persistence of {African} swine fever virus on porous and non-porous fomites at environmental temperatures},
	volume = {8},
	issn = {2055-5660},
	url = {https://porcinehealthmanagement.biomedcentral.com/articles/10.1186/s40813-022-00277-8},
	doi = {10.1186/s40813-022-00277-8},
	language = {en},
	number = {1},
	urldate = {2026-04-27},
	journal = {Porcine Health Management},
	author = {Nuanualsuwan, Suphachai and Songkasupa, Tapanut and Boonpornprasert, Prakit and Suwankitwat, Nutthakarn and Lohlamoh, Walaiporn and Nuengjamnong, Chackrit},
	month = dec,
	year = {2022},
	pages = {34},
}

@article{dorjee_network_2013,
	title = {Network analysis of swine shipments in {Ontario}, {Canada}, to support disease spread modelling and risk-based disease management},
	volume = {112},
	issn = {01675877},
	url = {https://linkinghub.elsevier.com/retrieve/pii/S0167587713002079},
	doi = {10.1016/j.prevetmed.2013.06.008},
	language = {en},
	number = {1-2},
	urldate = {2026-04-27},
	journal = {Preventive Veterinary Medicine},
	author = {Dorjee, S. and Revie, C.W. and Poljak, Z. and McNab, W.B. and Sanchez, J.},
	month = oct,
	year = {2013},
	pages = {118--127},
}

@book{usda_foot-and-mouth_2020,
	title = {Foot-and-mouth disease: {The} red book},
	url = {https://www.aphis.usda.gov/sites/default/files/fmd_responseplan.pdf},
	author = {USDA},
	year = {2020},
}

@article{osemeke_economic_2025,
	title = {Economic impact of productivity losses attributable to porcine reproductive and respiratory syndrome virus in {United} {States} pork production, 2016–2020},
	volume = {244},
	issn = {01675877},
	url = {https://linkinghub.elsevier.com/retrieve/pii/S0167587725002120},
	doi = {10.1016/j.prevetmed.2025.106627},
	language = {en},
	urldate = {2025-08-14},
	journal = {Preventive Veterinary Medicine},
	author = {Osemeke, Onyekachukwu and Silva, Gustavo S. and Corzo, Cesar A. and Kikuti, Mariana and Vadnais, Sarah and Yue, Xiaomei and Linhares, Daniel and Holtkamp, Derald},
	month = nov,
	year = {2025},
	pages = {106627},
}

@article{cardenas_spatio-temporal_2021,
	title = {Spatio-temporal network analysis of pig trade to inform the design of risk-based disease surveillance},
	volume = {189},
	issn = {01675877},
	url = {https://linkinghub.elsevier.com/retrieve/pii/S0167587721000581},
	doi = {10.1016/j.prevetmed.2021.105314},
	language = {en},
	urldate = {2025-10-06},
	journal = {Preventive Veterinary Medicine},
	author = {Cardenas, Nicolas Cespedes and VanderWaal, Kimberly and Veloso, Flávio Pereira and Galvis, Jason Onell Ardila and Amaku, Marcos and Grisi-Filho, José H.H.},
	month = apr,
	year = {2021},
	pages = {105314},
}

\end{document}